# Feasible Entanglement Purification for Quantum Communication

Jian-Wei Pan, Christoph Simon, Časlav Brukner & Anton Zeilinger

Institut für Experimentalphysik, Universität Wien, Boltzmanngasse 5, 1090 Wien, Austria

The distribution of entangled states between distant locations will be essential for the future large scale realization of quantum communication schemes such as quantum cryptography[1,2] and quantum teleportation[3]. Because of the unavoidable noise in the quantum communication channel, the entanglement between two particles is more and more degraded the further they propagate. Entanglement purification[4–7] is thus essential to distill highly entangled states from less entangled ones. Existing general purification protocols[4–6] are based on the quantum controlled-NOT (CNOT) or similar quantum logic operations, which are very difficult to implement experimentally. Present realizations of CNOT gates are much too imperfect to be useful for long-distance quantum communication[8]. Here we present a feasible scheme for the entanglement purification of general mixed entangled states, which does not require any CNOT operations, but only simple linear optical elements. Since the perfection of such elements is very high, the local operations necessary for purification can be performed with the required precision. Our procedure is within the reach of current technology and should significantly simplify the implementation of long-distance quantum communication.

Within the new fledgling field of quantum information[9], quantum communication has recently received much experimental attention. In the past years, entangled states have been used to experimentally demonstrate both dense coding[10–11] and teleportation[12–15]. Very recently, entanglement-assisted quantum cryptography has also been implemented using entangled photons[16–18]. While these schemes are realizable for moderate distances (up to a few kilometers in the case of cryptography), one is faced with serious problems, if one wants to go beyond this distance scale. One of the problems is the absorption of photons in the transmission channel. Since quantum states cannot be copied[19], the only way to solve this is by sending large numbers of photons. Nevertheless, photons still seem to be the best quantum information carriers over long distances.

Another problem is that typically the quality of the entangled states will decrease exponentially with the channel length. However, all the above protocols require that two distant parties, usually called Alice and Bob, share entangled pairs of high quality. Fortunately, various entanglement purification schemes have been suggested[4–7], by which Alice and Bob can generate a certain number of almost perfectly entangled pairs out of a larger number of less entangled pairs by performing local operations and using classical communication between the parties. This is important because the precision of both local operations and classical communication are independent of the imperfection of the quantum channel.

To show how entanglement purification works, the scheme introduced by Bennett et al.[4] is illustrated in Fig.1. From a practical point of view, the most important drawback of this scheme is that it requires the CNOT operation. In fact, the same is true for all known purification schemes working for general mixed entangled states. Although certain quantum logic gates have been experimentally demonstrated in physical systems such as ion-traps[20] and high-finesse microwave cavities[21], at present there is no implementation of CNOT gates that could realistically be used for purification in the context of long-distance quantum communication. This is due to the fact that for long-distance quantum communication purposes the probability of errors caused by the CNOT operation must not exceed the percent level[8], which is far outside the range of the present implementations. We will show how this problem can be overcome by presenting a general purification method that does not rely on the CNOT operation.

We consider qubits implemented as the polarization states of photons. We will denote the state of a horizontally polarized photon by $|H>$, and the state of a vertically polarized photon by $|V>$. In the usual quantum information terminology $|H>$ and $|V>$ correspond to $|0>$ and $|1>$.

Our purification scheme is based on a simple optical element, the polarizing beam splitter (PBS) (Fig. 2a). A PBS has two input modes 1 and 2, and two output modes 3 and 4. It acts like a mirror for vertically polarized light incident from either side, and like a transparent glass plate for horizontally polarized light. Consider the case where two photons are incident simultaneously, the first one along direction 1 and the second along direction 2. By selecting those events where there is one photon in each output mode one can perform a projection onto the subspace spanned by the two states $|H>|H>$ and $|V>|V>$, in which the two photons have equal polarization (see Fig. 2). We will show that this ability is sufficient to make entanglement purification possible. This feature of the PBS, which was first described in ref. 22, has been used in the observation of multi-photon entanglement[23–24], and also in a recent proposal for spin-flip-error correction in quantum communication[25].

Let us start the explanation of our purification scheme (shown in Fig. 3) by discussing a specific example. Suppose that Alice and Bob would like to share photon pairs in the specific maximally entangled state

$$|\Phi^+>_{ab} = \frac{1}{\sqrt{2}}(|H>_a|H>_b + |V>_a|V>_b) \tag{1}$$

where the photons at Alice's and Bob's locations are denoted by $a$ and $b$, respectively. Further suppose that before purification the pairs they share are all in mixed states

$$\rho_{ab} = F|\Phi^+>_{ab}<\Phi^+| + (1-F)|\Psi^+>_{ab}<\Psi^+| \qquad (2)$$

where $|\Psi^+>_{ab} = \frac{1}{\sqrt{2}}(|H>_a|V>_b + |V>_a|H>_b)$, i.e. there is an admixture of the unwanted state $|\Psi^+>_{ab}$. Note that in the state $|\Phi^+>_{ab}$ the two photons have equal polarization while in the state $|\Psi^+>_{ab}$ they have opposite polarization (for measurements in the *H/V* basis). Here we assume the special form (2) for simplicity only. The generality of our scheme will be proved later on.

Our proposed scheme is rather analogous to the scheme of Bennett et al.[4]. We also proceed by operating on two pairs at the same time. One can see that the main difference between Fig. 3 and Fig. 1 is that now each CNOT gate is replaced by a PBS. An essential step in our purification scheme is to select those cases where there is exactly one photon in each of the four spatial output modes, which we will refer to as "four-mode cases". As explained above, this corresponds to a projection onto the subspace where the two photons at the same experimental station (Alice's or Bob's) have equal polarization. Note that the polarizations at the two stations do not have to be the same.

From Eq. (2) it follows that the original state of the two pairs can be seen as a probabilistic mixture of four pure states: with a probability of $F^2$, pairs 1 and 2 are in the state $|\Phi^+>_{a1b1} \cdot |\Phi^+>_{a2b2}$, with equal probabilities of $F(1-F)$ in the states $|\Phi^+>_{a1b1} \cdot |\Psi^+>_{a2b2}$ and $|\Psi^+>_{a1b1} \cdot |\Phi^+>_{a2b2}$, and with a probability of $(1-F)^2$ in $|\Psi^+>_{a1b1} \cdot |\Psi^+>_{a2b2}$.

It is not hard to convince oneself that the cross combinations $|\Phi^+>_{a1b1} \cdot |\Psi^+>_{a2b2}$ and $|\Psi^+>_{a1b1} \cdot |\Phi^+>_{a2b2}$ never lead to four-mode cases. This is because, as mentioned above, the two entangled photons have equal polarization in the state $|\Phi^+>_{ab}$ while they have opposite polarization in the state $|\Psi^+>_{ab}$. Therefore, if the polarizations on Alice's side are equal, the

polarizations on Bob's side must be opposite, and vice versa. Thus, by selecting only four-mode cases one can eliminate the contribution of the cross terms. This is the basic principle of our purification method.

Further consider the two remaining combinations $|\Phi^+>_{a1b1} \cdot |\Phi^+>_{a2b2}$ and $|\Psi^+>_{a1b1} \cdot |\Psi^+>_{a2b2}$. Let us first discuss the $|\Phi^+>_{a1b1} \cdot |\Phi^+>_{a2b2}$ case. Selecting four-mode cases behind the two PBS projects the state $|\Phi^+>_{a1b1} \cdot |\Phi^+>_{a2b2} = \frac{1}{2}(|H>_{a1}|H>_{b1}+|V>_{a1}|V>_{b1}) \cdot (|H>_{a2}|H>_{b2}+|V>_{a2}|V>_{b2})$ into the non-normalized state

$$\frac{1}{2}(|H>_{a3}|H>_{a4}|H>_{b3}|H>_{b4}+|V>_{a3}|V>_{a4}|V>_{b3}|V>_{b4}) \qquad (3)$$

which is a maximally entangled four-particle state. This shows that the probability for a four-mode case is 50%. Alice and Bob can then generate maximal two-photon entanglement between the output modes $a3$ and $b3$ out of the four-photon entanglement by performing polarization measurements on each of the two photons at $a4$ and $b4$ in the $+/-$ basis and comparing their results, where $|+>=\frac{1}{\sqrt{2}}(|H>+|V>)$ and $|->=\frac{1}{\sqrt{2}}(|H>-|V>)$. If the measurement results at $a4$ and $b4$ are the same, i.e. $|+>|+>$ or $|->|->$, then the remaining two photons at $a3$ and $b3$ are left in the state $|\Phi^+>_{a3b3}$. If the results are opposite, i.e. $|+>|->$ or $|->|+>$, then the remaining two photons are left in the state $|\Phi^->_{a3b3} = \frac{1}{\sqrt{2}}(|H>_{a3}|H>_{b3}-|V>_{a3}|V>_{b3})$. In the second case either Alice or Bob could simply perform a local phase flip operation on his or her remaining photon to convert the state $|\Phi^->_{a3b3}$ back into $|\Phi^+>_{a3b3}$. Therefore in the $|\Phi^+>_{a1b1} \cdot |\Phi^+>_{a2b2}$ case Alice and Bob will get the state $|\Phi^+>_{a3b3}$ whenever there is exactly one photon in each output mode, i.e. with the probability 50%.

In the $|\Psi^+>_{a1b1} \cdot |\Psi^+>_{a2b2}$ case, following the same procedure Alice and Bob will project the remaining two photons $a3$ and $b3$ into the state $|\Psi^+>_{a3b3}$ with a probability of 50%. Since the probabilities to have a $|\Phi^+>_{a1b1} \cdot |\Phi^+>_{a2b2}$ and a $|\Psi^+>_{a1b1} \cdot |\Psi^+>_{a2b2}$ incident are $F^2$ and $(1-F)^2$ respectively, after performing the purification procedure Alice and Bob will finally obtain the state $|\Phi^+>_{a3b3}$ with a probability of $F^2/2$, and the state $|\Psi^+>_{a3b3}$ with a probability of $(1-F)^2/2$. By applying our purification procedure (selection of four-mode cases, measurements in modes a4 and b4 in the +/- basis, and local operations conditional on the results) they can thus create a new ensemble described by the density operator

$$\rho'_{ab} = F'|\Phi^+>_{ab}<\Phi^+| + (1-F')|\Psi^+>_{ab}<\Psi^+| \qquad (4)$$

with a larger fraction $F' = \frac{F^2}{F^2+(1-F)^2} > F$ (for $F > \frac{1}{2}$) of pairs in the desired state $|\Phi^+>_{ab}$ than before the purification. This concludes our discussion of purification for states of the form (2). To show that our scheme also works for general input states, let us analyze the relation between our scheme and the scheme of ref. 4 in more detail.

There is a close formal correspondence between our purification scheme and the scheme by Bennett et al. The CNOT gates in the ref. 4 serve exactly the same purpose as the polarizing beamsplitters in our procedure: they are used by Alice and Bob to determine whether the states of the two qubits at their respective locations are equal or opposite. This can be seen in the following way. From the logic table of the CNOT operation: $00 \to 00$, $01 \to 01, 10 \to 11$ and $11 \to 10$, it follows that after the operation the target particle is in state 0 if originally the two particles were in equal states, and it is in state 1 if they were in opposite states. In the protocol of ref. 4, after the CNOT operations on both sides Alice and Bob measure their target particles and keep the source pair if the target particles are both in the state 0 or both in 1. This means that the source pair is kept in two cases, namely, (1) when the two particles on Alice's side were in equal states, and the two particles on Bob's side were also in equal states (this corresponds to the case where both targets are in 0), and (2) when the two particles on Alice side were in opposite states, and the two particles on Bob side were also in opposite states (this corresponds to the case where both targets are in 1). In these two cases the state of the source pair will have higher entanglement than before.

The selection of four-mode cases behind the two PBS in our scheme exactly corresponds to the above case (1): coincidences behind a PBS imply that the polarizations of the two incoming photons were equal. One can show that there is full formal equivalence between the two schemes, apart from one fact: in our scheme we cannot make use of the above case (2). This means that for identical inputs the two procedures will lead to identical outputs in the case of success, but the success probability in our case will only be half as large as in the case of ref. 4. The formal equivalence of the two schemes also implies that the threshold fidelity that is required in order for the purification to work is the same for both, namely $F=\frac{1}{2}$. Furthermore, it is now clear that general mixed input states can also be purified in our scheme by using the methods described in the ref. 4, i.e. by additional bilateral local operations on individual pairs. Note that using the same method one can also realize a variant of the scheme by Deutsch et al.[5], which is more efficient than the scheme of ref. 4. Again the success probability would be lower by a factor of 2. This is a drawback of our PBS-based scheme, it increases the number of required original entangled pairs. However, we would like to mention that e.g. for the quantum repeater protocol[8] this increase scales polynomially (and not exponentially) with the channel length.

We have described a full entanglement purification method that works for general mixed entangled states and does not require the CNOT operation, in contrast to the previous schemes. With the techniques developed in experiments on quantum teleportation[12,13] and multi-photon entanglement[23-24], an immediate experimental verification of our purification protocol should be possible.

For true long-distance quantum communication one needs a scheme that allows the realization of many successive purification steps, perhaps most realistically following the quantum repeater protocol[8]. This imposes strict precision requirements on the local operations. Using the techniques referred to above, error probabilities on the percent level can be achieved. In particular the precision of linear optical elements such as polarizing beam splitters can be extremely high with errors on the order of a tenth of a percent. Furthermore, it is clear that for more elaborate purification protocols involving many steps at different locations a high-intensity source of entangled photon pairs is necessary. Currently the production rates in the present type of experiments are still rather low since parametric down-conversion, which is the standard source for entangled photons, occurs with small probability only. However, substantial progress is to be expected, for example, through cavity-enhanced down-conversion[26]. Finally let us emphasize that there is an enormous difference in the experimental effort required between implementing the CNOT operation and overlapping two photons on a polarizing beamsplitter. We believe that the present proposal might be a key ingredient for the future realization of long-distance quantum communication.

**Acknowledgement**

We are very grateful to L.-M. Duan, H. Ritsch, T. Tyc, P. Zoller and M. Zukowski for useful discussions. This work was supported by the Austrian Science Foundation FWF, the Austrian Academy of Sciences and the TMR program of the European Union.

Correspondence and requests for materials should be addressed to A. Z. (or e-mail: Zeilinger-office@exp.univie.ac.at).

**Figures**

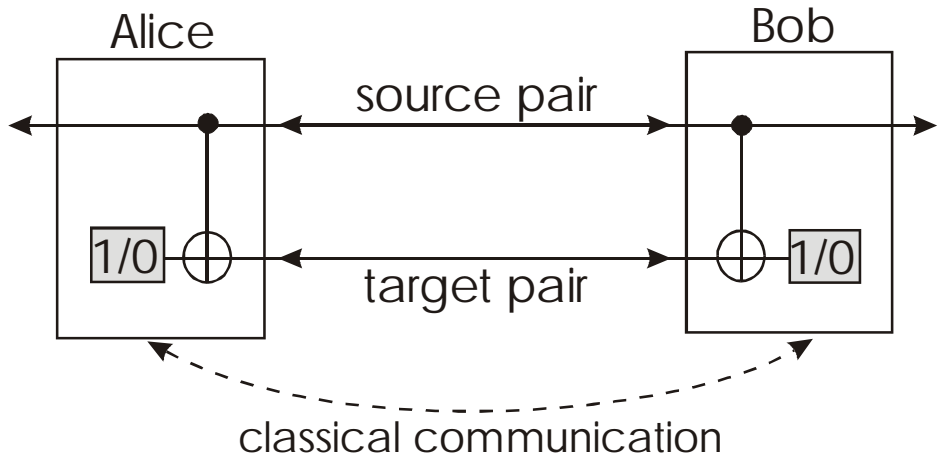

**Figure 1** Scheme showing the principle of entanglement purification after Bennett et al.[4]. One pair of higher entanglement is created starting from two less entangled pairs, where one member of each pair has been sent to Alice, and the other one to Bob. Both Alice and Bob perform a controlled-NOT (CNOT) operation on the two particles at their locations. Then they measure the particles belonging to the target pair in the computational basis (i.e., 0/1 basis) and compare the measured results via classical communication. It was shown in ref. 4 that if these results are the same then the remaining pair will have a higher degree of entanglement than the original two pairs, provided the quality of the original pairs was sufficiently high. Therefore in this case they keep the source pair. In the case of obtaining opposite results, they discard it. By repeating the same procedure, always starting from the pairs produced in the former purification step, it is possible to distill pairs of arbitrarily high entanglement quality. The higher the quality desired, the more original less entangled pairs are needed.

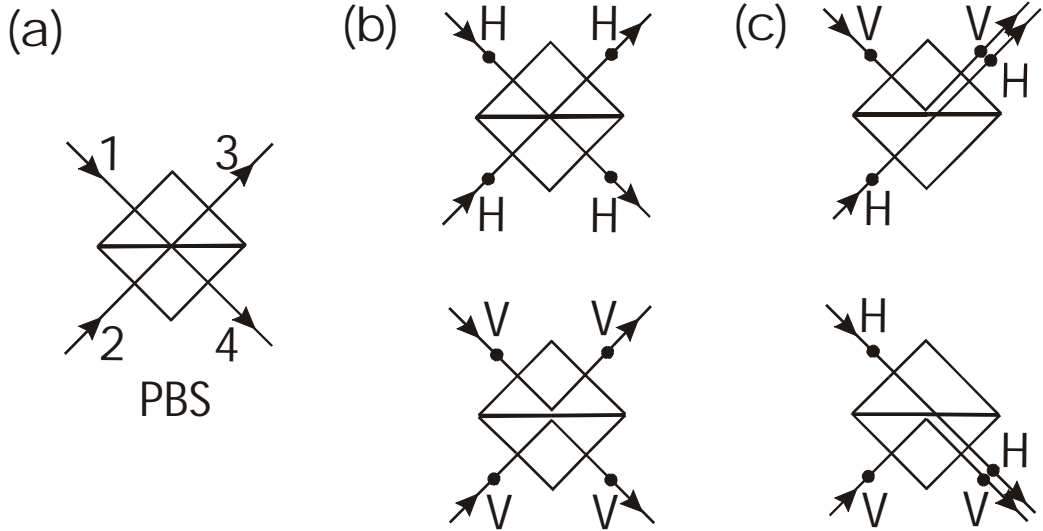

**Figure 2** Polarizing beam splitter (PBS) as polarization comparer. (a) The PBS transmits horizontal and reflects vertical polarization. This means, for example, that a vertically polarized photon incident along direction 1, denoted by $|V>_1$, goes out along direction 3, i.e. the state $|V>_1$ is transformed into $|V>_3$ by the action of the PBS. Similarly, $|H>_1$ goes to $|H>_4$, $|H>_2$ goes to $|H>_3$ and $|V>_2$ goes to $|V>_4$. (b) Now consider two photons incident simultaneously, one in each input mode with equal polarization, i.e. $|H>_1|H>_2$ or $|V>_1|V>_2$. Then they will always go out along different directions, i.e. there will be one photon in each of the two output modes. (c) On the other hand, if the two incident photons have opposite polarization, i.e. one is *V*- and the other one is *H*-polarized, then they will always go out along the same direction, i.e. there will be two photons in one of the two outputs and none in the other. In case (b), it is impossible in principle to determine whether the photons have been both transmitted or both reflected. This implies that finding one photon in each output mode corresponds to a projection onto the subspace spanned by $|H>_1|H>_2$ and $|V>_1|V>_2$.

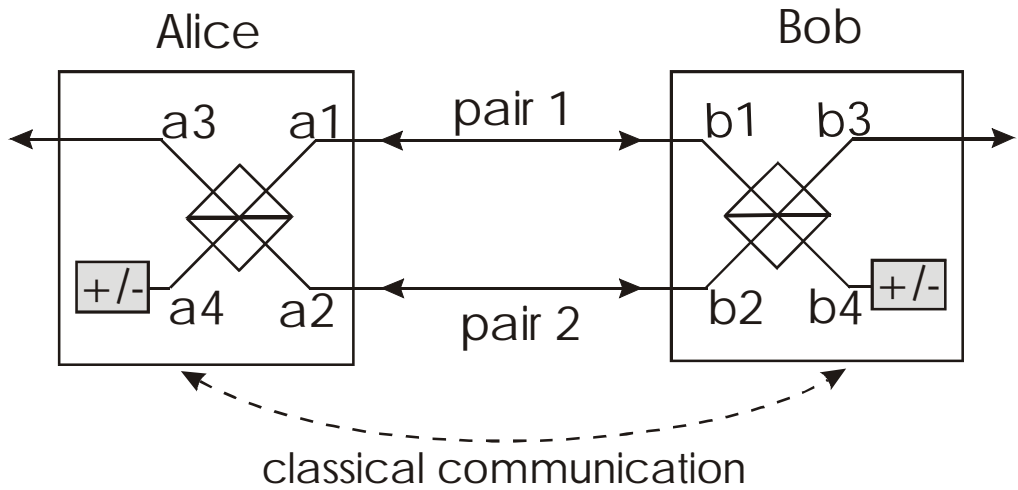

**Figure 3** Our purification scheme using polarizing beam splitters (PBS). Similarly to the scheme of Fig. 1, we start with two pairs shared by Alice and Bob. Both superimpose their photons on a PBS. They keep only those cases where there is exactly one photon in each output mode. Alice and Bob perform a polarization measurement in the 45 degree (+/-) basis in modes a4 and b4. Conditioned on the results Alice performs an operation on the photon in mode a3. After this procedure, the remaining pair in modes a3 and b3 will have higher degree of entanglement than the two original pairs. Note that it is not necessary to detect the photons in all four modes. It is sufficient to detect a single photon in each of the modes a4 and b4.